\documentstyle[preprint,aps,prl,epsbox]{revtex}
\tightenlines
\begin{document}
\title{Constraints on neutrino oscillation parameters from the measurement of day-night solar neutrino fluxes at Super-Kamiokande}

\draft
\maketitle

\begin{center}
\newcounter{foots}

The Super-Kamiokande Collaboration \\

Y.Fukuda$^a$, T.Hayakawa$^a$, E.Ichihara$^a$, K.Inoue$^a$,
K.Ishihara$^a$, H.Ishino$^a$, Y.Itow$^a$,
T.Kajita$^a$, J.Kameda$^a$, S.Kasuga$^a$
% ...

\footnotesize \it

$^a$Institute for Cosmic Ray Research, University of Tokyo, Tanashi,
Tokyo 188-8502, Japan\\
% ...
\end{center}

\begin{abstract}
A search for day-night variations in the solar neutrino flux 
resulting from neutrino oscillations has been carried out
using the 504 day sample of solar neutrino data obtained at Super-Kamiokande.
The absence of a significant day-night variation has set an absolute
flux independent 
exclusion region in the two neutrino oscillation parameter space. 
\end{abstract}


\narrowtext

As a real time solar neutrino experiment, Super-Kamiokande can 
perform a wide range of time modulation studies of the solar neutrino flux.
One motivation for these types of studies is an investigation
of neutrino oscillation hypotheses. All solar neutrino 
observations \cite{Cl,Kam3,SAGE,Gallex,SKflux}
have reported significantly lower fluxes than the expectations of standard 
solar models (SSMs)\cite{BP98,Bahcall95,SSMs}. 
This difference is commonly referred to as the solar neutrino problem.
Given the support of recent helioseismological 
observations\cite{helioseismology}, 
these SSMs look well-established and reliable. 
The difference between observations and predictions suggests some
neutrino properties beyond the standard model of elementary particles. 
The most popular solution to the solar neutrino problem
is neutrino oscillations, aided by matter
enhanced oscillations in the Sun\cite{MSW}. 
In some regions of the parameter space for neutrino oscillations,
matter enhanced oscillations within the Earth can lead to a
regeneration of the measured  neutrino flux passing through the Earth.
This regeneration would produce
a higher flux measured during night time relative to
day time measurements.
If such a day-night variation was observed, it would be strong evidence
for neutrino oscillations.  The amplitude of the day-night flux
variation would determine the neutrino oscillation
parameters, independent of the absolute flux uncertainties of the SSMs.

Super-Kamiokande (SK) started taking data in April, 1996. SK has already 
confirmed the deficit of solar neutrinos\cite{SKflux}. 
In this report, the total live time is increased to 503.8 days
(May $31^{st}$, 1996 through March $25^{th}$, 1998), 
and the total number of solar 
neutrino events found coming from the Sun is now $ 6823^{+148}_{-130}$ 
events above a threshold of 6.5 MeV
in total energy of the recoil electron.  Day-night (DN) variations are
investigated with this high-statistics solar neutrino data sample and
an updated flux value is presented.

The SK detector is located at the Kamioka Observatory, Institute for
Cosmic Ray Research, the University of Tokyo,
in Gifu Prefecture, Japan, 137.32 
degrees East longitude and 36.43 degrees North latitude.
Due to the latitude, the nadir of the Sun 
can range between $\pm 0.974$ in cosine (12.98 to 167.02 degrees) 
at the SK site. In this analysis, nadir ($\theta _{z}$) is defined as the
angle between the negative z-axis of the detector coordinate system and 
the direction to the Sun (solar neutrino direction),
where the cosine of the nadir is positive when the Sun is below
the horizon.  Solar neutrinos will penetrate different regions of
the Earth, that are related to the nadir of the Sun by the following:
mantle ($0 < \cos\theta_{z} < 0.838$), outer core 
($0.838 < \cos\theta_{z} < 0.981$) and inner 
core ($0.981< \cos\theta_{z} < 1$).
SK never sees solar neutrinos that pass through inner core of 
the Earth. However, the fact 
that the density of the outer core ($10\sim 12 g/cm^3$) is about double 
that of the mantle ($< 5.5 g/cm^3$)
may enhance the neutrino regeneration efficiency
of the Earth.  To look for effects associated with the core,
the night sample was divided into
5 data sets according to nadir of the Sun at the time of the neutrino event,
N1 ($ 0 < \cos \theta_{z} \leq 0.2$), 
N2 ($ 0.2 < \cos \theta_{z} \leq 0.4$) ... ,
and N5 ($ 0.8 < \cos \theta_{z} \leq 1.0 $).
In the cases of N1 to N4, neutrinos pass only through the mantle of the Earth.
N5 data contains the outer-core-penetrating neutrino sample.  In the N5 
period, the Sun has spent 80\% of its time in the outer core region.

The absolute energy scale of the SK detector has been calibrated 
precisely using a linear accelerator system (LINAC) 
at the detector\cite{linac}.  The LINAC injects
electrons downward in direction, at several fixed points in the 
volume, and at certain times. 
Uniformity of the detector response in each nadir direction is important,
since the data is divided according to the nadir angle.
The uniformity in azimuthal angle is also important since the nadir and 
azimuthal angles of the Sun are correlated.
In addition, N5 data can only be taken 
during the winter time, so long term stability
of the detector must be monitored.
The precise LINAC calibration has to be extrapolated to the entire volume,
in all directions, and at all times.
Decay of spallation products induced by cosmic ray 
muons are used for this purpose. 
Spallation events accumulate in the data as a major background 
in the solar neutrino measurements ($\sim$ 600 events/day above 
6.5 MeV threshold in a 22.5 ktons fiducial volume).
They are distributed uniformly in volume, direction, and 
time, just as solar neutrinos. Their beta decay energy spectra are 
distributed in the relevant energy region for $^8B$ neutrinos,
making this background a good calibration and stability monitor. 
The spallation event sample was divided
into subsets in time and direction (nadir and azimuthal angle) 
and the relative energy difference was
found by comparing the spectral shape of a subset with that of
the whole spallation sample as a reference spectrum.
The magnitude of spread of the relative energy differencies is 
consistent with the expected statistical distribution of subsets and no 
systematic biases are seen in the test variables.  A conservative
systematic error for relative energy scale for each N1 to N5 data set is 0.5\%.
Due to the steepness of the recoil electron spectrum
near the analysis threshold of 6.5 MeV, this 
small scale error is amplified and
the relative flux error of each data subset (N1-N5) in the uniformity of
the energy scale is estimated to be $^{+1.2}_{-1.1}\%$. 
Since the statistical errors of the flux values are much larger than 
this scale error and the spread of the relative energy scales was consistent 
with a statistical distribution, we treated this scale error
as uncorrelated among subsets in the following analysis.

In order to be independent of the 
absolute flux values of $^8B$ solar neutrinos in SSMs,
only the relative difference of DN flux values were used in the oscillation 
analysis.
In this case, many of the systematic errors involving detector response
cancel. The remaining systematic errors, in addition to
the relative energy scale error, are estimated 
to be: $\pm 0.1\% $ (data reduction), $\pm 0.4\%$ (background subtraction),
$\pm 0.1\%$ (live time calculation), and $<0.1\%$ (trigger efficiency) 
for each night subset.

\begin{figure}[htb]
\begin{tabbing}
\= \kill
\hspace{0mm}
\mbox{\epsfile{file=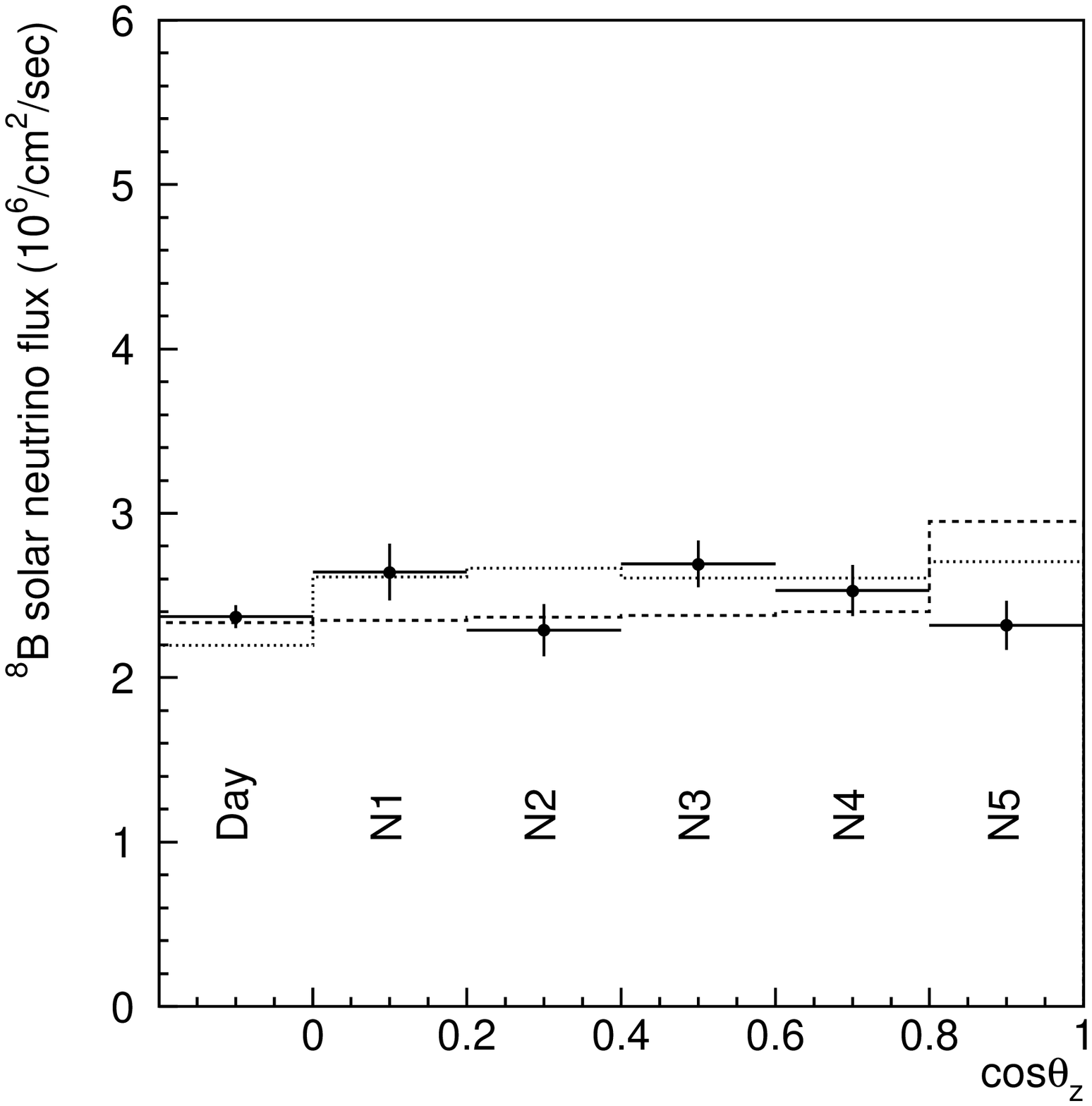,height=15cm}}
\> \hspace{80mm} \mbox{\epsfile{file=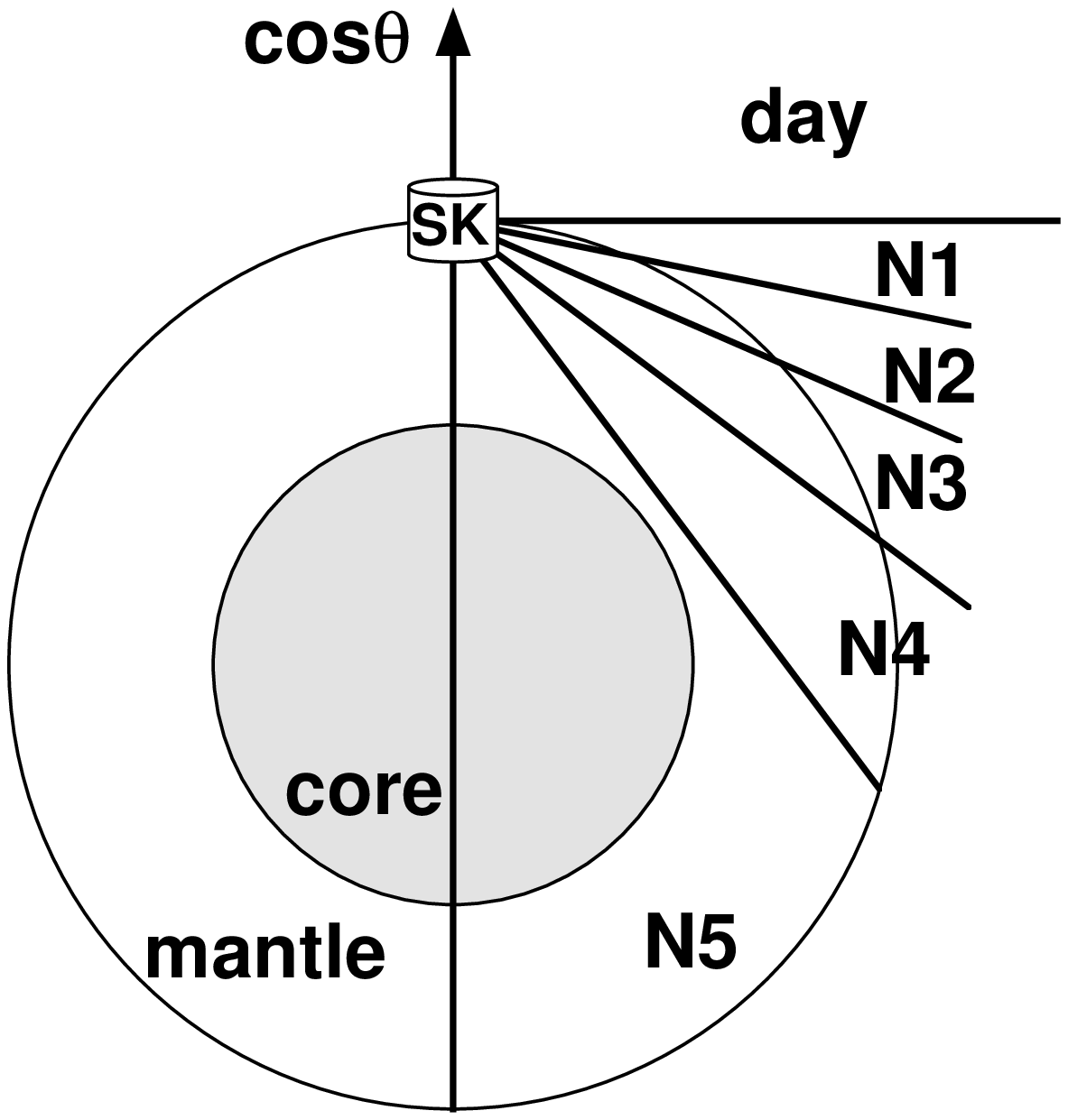,height=13cm}}
\end{tabbing}
\caption[]{ Measured day/night solar neutrino fluxes as a function of
the nadir of the Sun. Error bars represent statistical errors only.
Night data is divided into 5 bins. Dotted histogram is the expected variation 
of a typical large angle solution and dashed histogram is that of a typical
small angle solution.}
\label{fig:dn}
\end{figure}

Figure~\ref{fig:dn} shows the measured DN fluxes from 503.8 days of SK data.
The flux is normalized to a value at one astronomical unit (AU) by a correction
for the eccentricity of the Earth's orbit.  
Numerical values of the corrected fluxes are listed in 
Table~\ref{tab:results} with relative systematic errors. The total flux
updated for 503.8 days of data is also listed in the table with an absolute
systematic error same as previously 
presented\cite{SKflux}. Obtained DN asymmetries
are as follows:
$$ \frac{N}{D}-1 = 0.047 \pm 0.042(stat) \pm 0.008(syst) $$
$$ \frac{N5}{\left \langle D, N1\cdot\cdot\cdot N4 \right \rangle}-1 = -0.055 \pm 0.063(stat) \pm 0.013(syst) $$
No significant DN variation nor N5 excess is seen in the data. 

The impact of these results was investigated within a two neutrino 
oscillation hypothesis for
$ \nu _e \rightarrow \nu _{\mu} $ or $ \nu _e \rightarrow \nu _{\tau}$ 
(active neutrinos), 
and $ \nu _e \rightarrow \nu _s$ (sterile neutrino).
A flux independent analysis was performed by treating the 
flux normalization factor $\alpha $ as a free parameter in the 
$\chi ^2$ definition.
$$  \chi^2_{DN}=\sum_{i=D,N1,\cdot\cdot\cdot,N5}\left\{\frac{\phi _i
-\alpha\times \phi^{osci}_i(\sin^22\theta, \Delta m^2)}{\sqrt{\sigma_i^2+\sigma_{sys,i}^2}}
\right\}^2 $$
where $\phi _i$ is the measured flux, 
$\phi^{osci}_i(\sin^22\theta, \Delta m^2)$
is the effective flux for a given set of oscillation parameters 
derived from the ratio of the expected number of events with and without 
oscillations.  $\sigma_i$ and $\sigma_{sys,i} $
are the statistical and systematic errors of i$^{th}$-bin listed in 
table~\ref{tab:results}.
The $\chi^2$ value for the case of no oscillations
is 7.4 with 5 degrees of freedom (dof), 
which corresponds to a 19\% probability.
The expected solar neutrino flux nadir dependencies
for a set of oscillation parameters were obtained 
by a numerical calculation using models 
for the neutrino production point\cite{Bahcall95}, 
electron density in the Sun\cite{Bahcall95}, electron density
in the Earth (PREM\cite{PREM}) and $^8B$ neutrino spectrum\cite{Bahcall96}.
Production points were integrated on a $0.01 R_{sun}$ grid in the plane
which contains the Sun and the Earth.
The electron density distribution in the Earth was calculated with 
charge-to-mass ratios (Z/A) of 0.468 for the core and 0.497 
for the mantle\cite{BKdaynight}.
Neutrino trajectories in the Earth were integrated
over 1000 directions with $0.001 \cos \theta_{z}$ 
steps, weighted by SK live time. 
Assuming neutrino incoherence at the Earth, 
the electron neutrino survival probability at the detector, $P_{SE}$, was
obtained from independent calculations of 
$P_{1,2}(\Delta m^2/E,\sin^22\theta,{\bf r}_0)$ and $P_{1e,2e}(\Delta m^2/E,\sin^22\theta,\cos\theta_{z})$ using
$$ P_{SE} = P_1P_{1e} + P_2P_{2e} = (1-P_2)(1-P_{2e}) + P_2P_{2e}$$
where $P_1, P_2$ are the probabilities to be $\nu_1, \nu_2$ 
at the surface of the Sun
and ${\bf r}_0$ is the production point of neutrinos in the Sun.
$P_{1e}, P_{2e}$ are the probabilities to be detected as a $\nu_e$ at SK 
if the neutrinos arrive as $\nu_1, \nu_2$,
taking into account any possible regeneration in the Earth.
Representative flux variations obtained from these 
calculations (for active neutrinos) are shown 
in figure~\ref{fig:dn}
for a typical set of parameters at a large mixing angle solution 
($\sin^22\theta=0.56, \Delta m^2=1.2\times 10^{-5}eV^2$)
and for a set at a typical small mixing angle solution 
($\sin^22\theta=0.01, \Delta m^2=6.3\times 10^{-6}eV^2$).

\begin{figure}[htb]
\begin{tabbing}
\= \kill
\hspace{0mm}
\hspace{0mm} \mbox{\epsfile{file=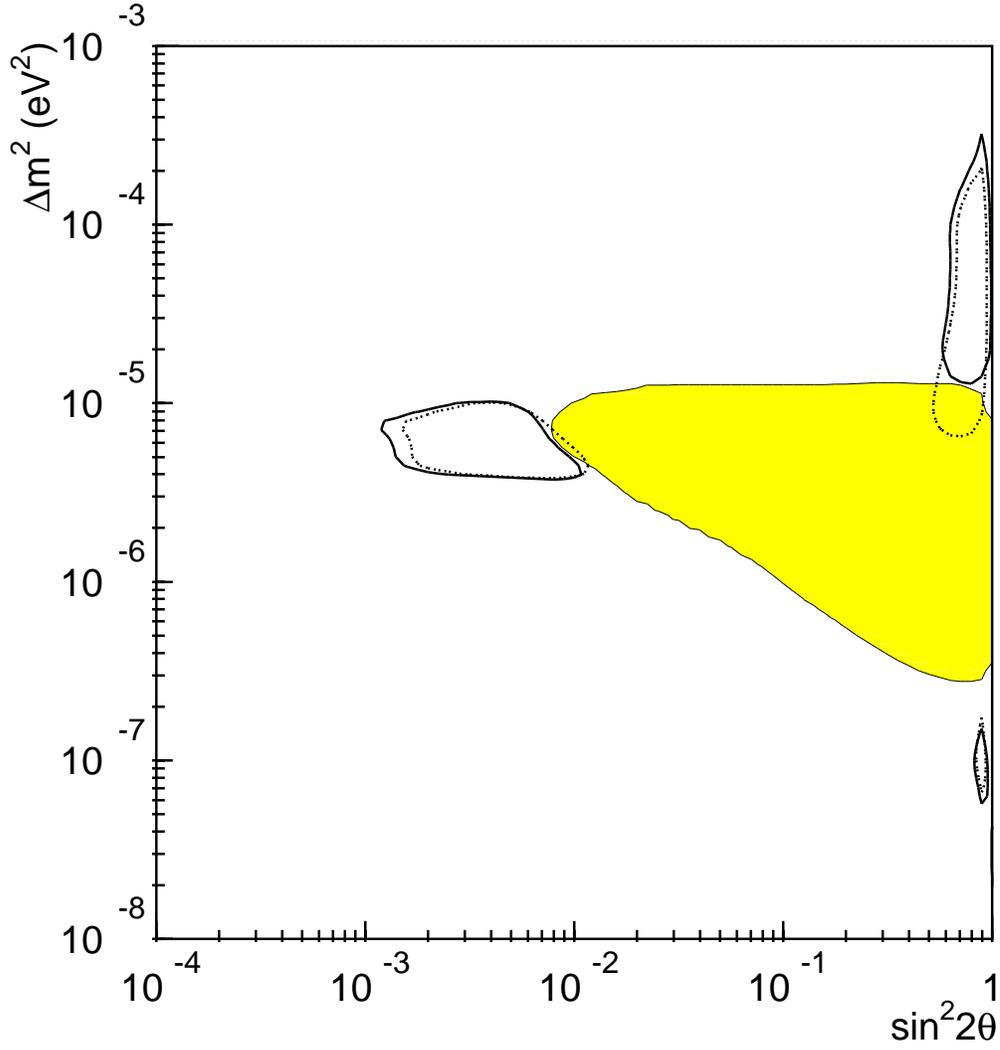,height=15cm}}
\end{tabbing}
\caption[]{ Flux independent exclusion region by SK day/night variation for
$\nu_e \rightarrow \nu_{\mu,\tau}$ oscillations. Exclusion probabilities
larger than 99\% are shown in the shaded area. 
Regions inside of the dotted lines are
allowed at the 99\% C.L. from the combined 
rate analysis of Homestake, SAGE, Gallex
and SK-flux in comparison with the BP98 SSM\cite{BP98}. 
Regions inside of the thick
solid lines are allowed at the 99\% C.L. from the combined rate analysis of 
the rates and the SK D/N variation.}
\label{fig:contour_active}
\end{figure}

\begin{figure}[htb]
\begin{tabbing}
\= \kill
\hspace{0mm}
\hspace{0mm} \mbox{\epsfile{file=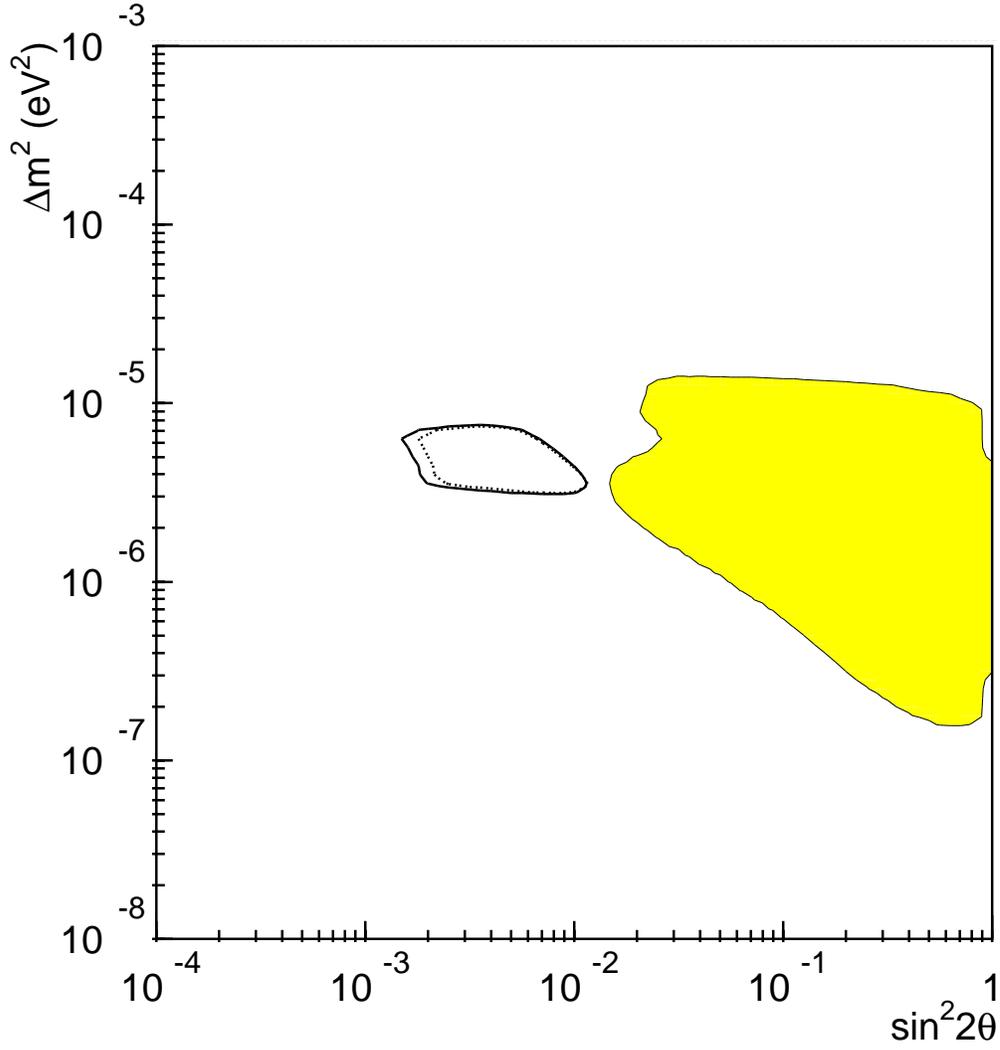,height=15cm}}
\end{tabbing}
\caption[]{ Flux independent exclusion region by SK day/night variation 
for $\nu_e \rightarrow \nu_s$ oscillations.  Regions are defined as in
Figure~\ref{fig:contour_active}.}
\label{fig:contour_sterile}
\end{figure}

The expected solar neutrino DN flux was calculated for points
in the parameter space ($ 10^{-4}\leq \sin^22\theta \leq 1,
10^{-8} \leq \Delta m^2 \leq 10^{-3} eV^2$) and compared to the
measured values.
Minimum $\chi^2$ values of 5.2 were found 
at ($\sin^22\theta=3.2\times 10^{-2}, \Delta m^2=1.6\times 10^{-6}eV^2$)
for the $\nu_e \rightarrow \nu_{\mu,\tau}$ case and 4.8 
at ($\sin^22\theta=0.25, \Delta m^2=1.8\times 10^{-7}eV^2$)
for the $\nu_e \rightarrow \nu_s$ case.
These sets of parameters can not explain the 
deficits seen in the other experiments, so
a hypothesis tests based on $\chi^2_{DN}$
was performed to obtain an exclusion 
region, instead of a method 
based on a distance from the minimum $\chi^2$.
Figures~\ref{fig:contour_active} and \ref{fig:contour_sterile} show
shaded regions where the exclusion probabilities are more than 99\% 
($\chi^2>15.09$ for 5 dof),
using the expected nadir dependencies 
for $\nu_e \rightarrow \nu_{\mu,\tau}$ and 
$\nu_e \rightarrow \nu_s$, respectively. 
The typical parameters shown in figure~\ref{fig:dn} are excluded.
Absolute flux information was not used in these calculations so the
excluded regions found are independent of the $^8B$ neutrino flux.

A combined rate analysis was performed using data from the
4 different solar neutrino experiments (Cl, SAGE, Gallex and SK) 
with the assumption of the absolute fluxes given in \cite{BP98}.
The analysis follows the method given in \cite{global} with
updated theoretical uncertainties ($^{37}Cl$ cross section \cite{cl_error}, 
$^{71}Ga$ cross section\cite{ga_error} and diffusion\cite{Bahcall95}) 
and the latest experimental results\cite{latest}.
In order to see the impact of the DN information obtained at SK,
the rate analysis  was performed with and without the 
DN information from SK.
Results from this combined rate analysis are also shown in
figures~\ref{fig:contour_active} and \ref{fig:contour_sterile}.
Regions inside the solid and dotted lines show the allowed region 
at the 99\% C.L. ($\chi^2<\chi^2_{min}+9.21$) 
with and without DN information, respectively.

In summary, SK observes no DN variation in the $^8B$
solar neutrino flux in 504 days of
data and sees no significant zenith angle variations. This result sets
an exclusion region in the oscillation parameter space where regeneration of
$\nu_e$ would be expected. That exclusion region constrains the MSW solutions to
larger mass difference (large angle solution) or to smaller mixing angle 
(small angle solution). \\

We gratefully acknowledge the cooperation of the Kamioka Mining and Smelting 
Company. This work was partly supported by the Japanese Ministry of Education,
Science, Sports and Culture and the U.S. Department of Energy.

\begin{table}[htb]
\caption{Day/Night fluxes obtained from 503.8 days of SK data. Flux values are
normalized to 1 AU. Quoted systematic error for the "All" data is 
an absolute error but others are relative errors. }
\label{tab:results}
\begin{tabular}{cclcr}
         Data set & Nadir of the Sun &  Flux ($10^6/cm^2/sec$) &Syst. error& Live Time \\
\hline
\hline
Day & $-1 \leq \cos \theta _{z} \leq 0$ & $ 2.369^{+0.072}_{-0.068}(stat) $& $^{+0.6}_{-0.5}\%$ & 242.2 days \\
Night & $0 < \cos \theta _{z} \leq 1$ & $ 2.481^{+0.068}_{-0.065}(stat) $&$^{+0.6}_{-0.5}\%$ & 261.6 days \\
N1  & $ 0 < \cos \theta _{z} \leq 0.2$ & $ 2.640^{+0.170}_{-0.167}(stat) $&$^{+1.3}_{-1.2}\%$ & 43.6 days \\
N2  & $0.2 < cos \theta _{z} \leq 0.4$ & $ 2.289^{+0.157}_{-0.154}(stat) $&$^{+1.3}_{-1.2}\%$ & 48.6 days \\
N3  & $0.4 < cos \theta _{z} \leq 0.6$ & $ 2.689^{+0.144}_{-0.134}(stat) $&$^{+1.3}_{-1.2}\%$ & 64.3 days \\ 
N4  & $0.6 < cos \theta _{z} \leq 0.8$ & $ 2.526^{+0.156}_{-0.146}(stat) $&$^{+1.3}_{-1.2}\%$ & 52.5 days \\
N5  & $0.8 < cos \theta _{z} \leq 1$ & $ 2.318^{+0.147}_{-0.144}(stat) $&$^{+1.3}_{-1.2}\%$ & 52.6 days \\
\hline
All & $-1 \leq cos \theta _{z} \leq 1$ & \multicolumn{2}{l}{$ 2.436^{+0.053}_{-0.047}(stat) ^{+0.085}_{-0.071}(syst)$} & 503.8 days \\
\end{tabular}
\end{table}

\end{document}